\documentclass{article}
\usepackage[utf8]{inputenc} 
\usepackage[T1]{fontenc}    
\usepackage{nicefrac}       
\usepackage{microtype}      
\usepackage{graphicx}
\usepackage{upgreek}
\usepackage{amsmath}
\usepackage{mathtools}
\usepackage[boxruled,algoruled]{algorithm2e}
\usepackage{float}
\usepackage{bigints}
\usepackage{xcolor}
\usepackage{framed}
\usepackage[htt]{hyphenat}
\usepackage{url}
\usepackage{makecell}
\usepackage{tabularx}
\usepackage{xr-hyper}
\usepackage{arxiv}
\usepackage[printwatermark]{xwatermark}

\newcommand{\micron}{$\upmu$m}

\newcommand{\vecs}{\boldsymbol{s}}

\newcommand{\vecy}{\boldsymbol{y}}

\newcommand{\mataa}{\boldsymbol{A}}

\newcommand{\norm}[1]{\left\lVert#1\right\rVert}

\newcommand{\gammaexcl}{\gamma_{\text{excl}}}

\newcommand{\ionent}{I_1^0}
\newcommand{\itwont}{I_2^0}
\newcommand{\ionetwont}{I_{1/2}^0}
\newcommand{\zdof}{z_{\text{DoF}}}
\newcommand{\deltares}{\delta_{t}}

\graphicspath{{figures/}}

\title{Using a modified double deep image prior for crosstalk mitigation in multislice ptychography}

\author{
 Ming Du \\
  Advanced Photon Source \\
  Argonne National Laboratory \\
  Lemont, Illinois 60439, USA \\
  mingdu@anl.gov \\
  \And
 Xiaojing Huang \\
  National Synchrotron Light Source II \\
  Brookhaven National Laboratory \\
  Upton, New York 11973, USA \\
  \And
 Chris Jacobsen \\
  Advanced Photon Source \\
  Argonne National Laboratory, Lemont, Illinois 60439, USA \\
  \{Department of Physics \&{} Astronomy, Chemistry of Life Processes Institute\} \\
  Northwestern University \\
  Evanston, Illinois 60208, USA \\
  cjacobsen@anl.gov
}

\date{}

\begin{document}

\maketitle

\section*{Abstract}

Multislice ptychography is a high-resolution microscopy technique used
to image multiple separate axial planes using a single illumination
direction.  However, multislice ptychography reconstructions are often degraded
by crosstalk, where some features on one plane erroneously contribute
to the reconstructed image of another plane.
Here, we demonstrate the use of a modified ``double deep image prior'' (DDIP) architecture in mitigating
crosstalk artifacts in multislice ptychography. Utilizing the tendency
of generative neural networks to produce natural images, a modified DDIP method yielded good results on experimental data.
For one of the datasets, we show that using DDIP could remove the need of using additional
experimental data, such as from x-ray fluorescence, to suppress the crosstalk. Our method may help
x-ray multislice ptychography work for more general experimental scenarios.

\section{Introduction}

In ptychography, a spatially-limited coherent probe is scanned across
multiple transverse positions; the collection of far-field diffraction
patterns are then used to reconstruct the complex optical
transmittance of a planar object \cite{faulkner_prl_2004}.  Multislice
ptychography \cite{maiden_josaa_2012, tsai_optexp_2016b} is an
extension of this approach for imaging multiple axial planes each separated
by a distance $\zdof$ greater than the depth of field (DoF) of
\cite{born_optics,gilles_optica_2018}
\begin{equation}
  \zdof = \frac{2}{0.61^{2}}\frac{\deltares^{2}}{\lambda} \simeq
    5.4 \frac{\deltares^{2}}{\lambda}
  \label{eqn:dof}
\end{equation}
where $\deltares$ is the transverse spatial resolution.  In multislice
ptychography, the probe illumination function at each probe position
is modulated by the first axial plane, after which Fresnel propagation
is used to bring it to the next plane, and so on until the far-field
diffraction intensity is obtained.

When the contrast of upstream planes is significant enough that the
first Born approximation is violated, the illumination of downstream
planes is significantly affected; if incorrectly accounted for in a
reconstruction algorithm, this can lead to crosstalk between the
images from these separate planes.  Even with low contrast objects, if
the axial separation between object planes is only a small multiple of
$\zdof$, Fresnel propagation alone may be insufficient to cleanly
reconstruct the two planes correctly. This can be seen in a 12 keV
x-ray multislice ptychography experiment where crosstalk was obseved
in $\deltares=9.2$ nm images of objects on two planes separated by 10
\micron{}, or 2.3 times $\zdof=4.4$ \micron{} in this case
\cite{ozturk_optica_2018}.
Given that hard x-ray microscopy is
well suited to imaging objects in this thickness range
\cite{du_ultramic_2018}, this limitation of multislice ptychography
becomes important to overcome.  Alternative approaches include
ptychographic tomography for objects that do not extend in depth
beyond $\zdof$ at any rotation angle \cite{dierolf_nature_2010}, or
multislice ptychographic tomography of thicker objects where propagation is used to
compensate for Fresnel diffraction blurring but images of separate
planes are not required
\cite{vandenbroek_prl_2012,kamilov_optica_2015,li_scirep_2018,gilles_optica_2018,du_sciadv_2020}; however,
both of these approaches require images obtained over multiple
object rotation angles.  The more extensive data collection required
for these tomographic approaches is not always feasible or desirable,
so it remains important to overcome crosstalk effects in
single-viewing-direction multislice ptychography of separate object
planes.

Many ptychographic beamlines at synchrotron light sources are equipped with both an area detector for recording
far-field coherent diffraction data, and an energy-dispersive detector
for recording x-ray fluorescence (XRF) signals in the same scan of the illumination
probe. Unlike ptychography, fluorescence imaging is an incoherent
process with a spatial resolution limited by the focusing optic used;
however, XRF can provide low spatial frequency information of a sample
with distinct distributions of chemical elements.  This approach has
been used to provide low-crosstalk reconstructions of an upstream
plane object consisting of an Au zone plate structure and a downstream
plane consisting of NiO particles mounted on a silicon nitride window
\cite{huang_actacrysta_2019}.  In this case, the Ni XRF image was used
to generate an initial guess of the object on the downstream plane, as
well as to subtract the spectrum of the NiO object's ``ghost image'' from an initial
reconstruction of the upstream plane, after which a multislice
ptychographic reconstruction was allowed to proceed.  The resulting
images (shown in Figs.~3(a) and (b) of \cite{huang_actacrysta_2019})
indeed show almost no crosstalk between the reconstructed images at
the two axial planes.

While the XRF-aided reconstruction has been shown to be effective, its
limitation is also apparent: if the chemical composition of objects on
the different axial planes is similar, then XRF can no longer provide
strict object separation.  Therefore, it is valuable to explore
alternative methods to suppress crosstalk without using XRF data. In
fact, the crosstalk separation problem resembles the well-known
problem of blind source separation (BSS) problem in signal processing
\cite{cao_ieeetrans_1996}. In the BSS problem, one begins with $N$ measurement
$\vecy = [y_{1}(t), y_{2}(t), \cdots, y_{N}(t)]$, where each measurement is
a linear superimposition of $M$ source signals
$\vecs = [s_{1}(t), s_{2}(t), \cdots, s_{M}(t)]$ with a unique set of
weighting factors $w_{n,m}$ so that one obtains measured data of $y_{n}(t) =
\sum_{m}^{M} w_{n,m} s_{m}(t)$.  The goal in this case is to solve the
linear system
\begin{equation}
  \vecy = \mataa\vecs
  \label{eqn:linear_equation}
\end{equation}
so as to obtain the source signals
$\vecs$.  The problem can be
overdetermined, underdetermined, or exactly determined depending on
the relative values of $M$ and $N$.  Separating out all $M$ sources
requires $N \geq M$. Obviously, for multislice ptychography, $N = M$,
which is a necessary condition for all ``clean'' slices to be solved
from phase retrieved slices containing crosstalk.

The complication for multislice ptychography is that the ghost
features are not a simple superimposition added onto an affected
slice, but rather a filtered verison of the real features after losing
information in certain spatial frequency bands.  For example, the ghost
particles in one axial plane of Fig.~4(a) of \cite{ozturk_optica_2018}
appear like a low-pass filtered version of features in the other axial
plane. This band loss has to be taken into account before separating
the ghost features. Moreover, for a BSS problem to be solved
successfully, the rows of $\mataa$ in Eq.~\ref{eqn:linear_equation}
should be linearly independent. In the case of multislice
ptychography, that requires sufficient differentiation between real
and ghost features in the axial slices. When the slice spacing is
large, this condition is usually easy to satisfy. However, if the
slice separation is too small, the weak probe variation between
adjacent slices makes them hard to be cleanly reconstructed when
starting from a random guess, since this can yield retrieved slices
that are too similar to each other. Under this scenario, we may relax
our constraint and allow the use of XRF data to assist with the
initial phase retrieval. However, it turns out that even with good
initial guesses aided by XRF, one is still unable to fully eliminate
inter-slice crosstalk without very a careful search of reconstruction
parameters and reconstruction algorithms. For example, to obtain
Fig.~3(a) and (b) of \cite{huang_actacrysta_2019}, many efforts were
made to optimize the algorithm and parameters. Before doing that, a
standard reconstruction yielded images with considerable crosstalk as
shown in Fig.~4(a) of \cite{huang_actacrysta_2019} and as will be
shown again below.  We demonstrate here that crosstalk can be greatly
reduced, so that in both situations (large separation without using
XRF, and small separation with XRF), the crosstalk can be mitigated
using a neural network algorithm based on a ``double deep image
prior'', or ``double-DIP'' (DDIP).

In the deep image prior (DIP) approach \cite{ulyanov_arxiv_2018},
images in the forward model are generated from a generative neural
network, so that the network itself functions to provide prior
knowledge to the system.  This is because a deep neural network
prefers generating ``natural images'' with lower patch-wise entropy,
rather than those with higher patch-wise entropy \cite{gandelsman_arxiv_2018}.
In fact, the DIP is a type of ``untrained'' neural network, which
means that instead of using a large dataset to train a DIP, one uses
the original image in a specific image reconstruction task, and uses
the trained DIP only for this task itself. As such, a shallow network
suffices for a DIP since it only needs to ``learn'' from the image(s)
being processed, and the generalizabilty of a DIP-based algorithm is
not constrained by the training set.  This initial demonstration of
DIP \cite{ulyanov_arxiv_2018} used an encoder--decoder structure which
learns to map an input tensor to an image with the same spatial
dimension. By using an encoder--decoder network with skip connections
linking the encoder part and the decoder part, one can lead the
network to generate images with structure at multiple spatial scales,
thus better capturing the characteristics of natural images.

Building upon this initial work, it has been shown that the use of multiple DIP networks
can achieve improved outcomes on a series of layer decomposition
problems including image dehazing,
image segmentation, and transparency separation whose goal is to
separate out multiple individual natural images from blends of them \cite{gandelsman_arxiv_2018}.
All these tasks can be carried out using a similar architecture: if there are two layers to be separated, one can use
two DIP networks to generate two distinct images, and use a third DIP to generate either a mask or a constant blending ratio.
The architecture is thus named ``double-DIP'' (DDIP) after the two image-generating DIPs.
Using the generated images and the mask or ratio, one can synthesize a blended image, and train the networks to minimize a
loss function measuring the mismatch between the synthesized image and the original blended images.
The preference of DIPs to generate natural images
means that the local patches consisting the images they generate
usually have lower empirical entropy, which is an indication that
these images are more likely to be unblended ``single''
images. Additionally, prior work \cite{gandelsman_arxiv_2018} also included an exclusion term
in the loss function, which penalizes the correlation between the spatial gradients of the generated images. This further
suppresses the crosstalk in the output images.

Therefore, one can expect that the DDIP architecture can function effectively in the multislice ptychography crosstalk
separation problem. In view of the additional band loss complication of the ghost features, we modified the DDIP architecture
from the original design of \cite{gandelsman_arxiv_2018}. The network architecture will be introduced in more detail
in Section \ref{sec:methods}. In Section \ref{sec:results}, we will the show the results obtained using DDIP on two datasets,
each representing one of the slice spacing situations mentioned above.

\section{Methods}
\label{sec:methods}

\subsection{Algorithm}

\begin{figure}[htbp]
  \centering
  \includegraphics[width=0.7\textwidth]{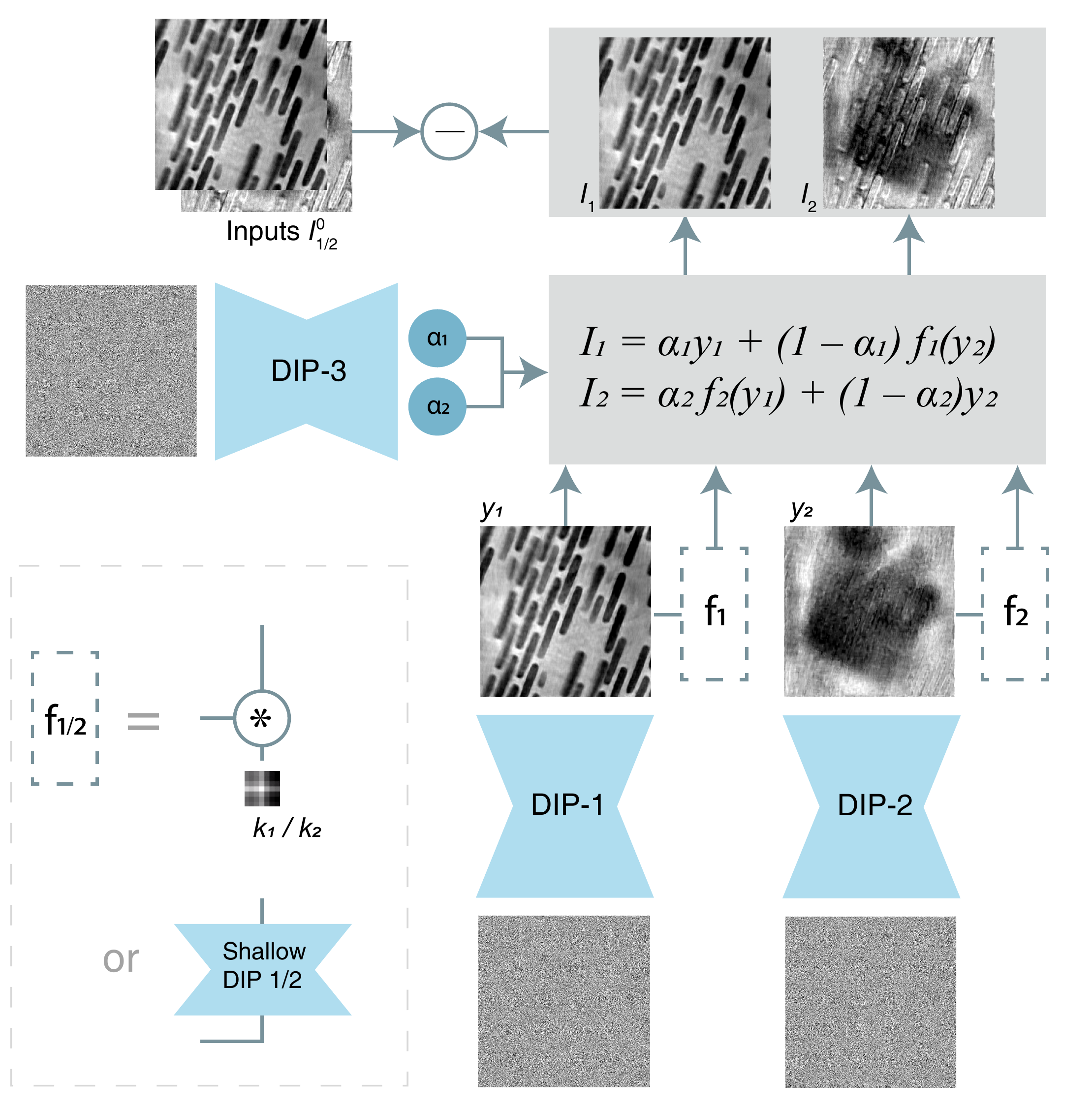}
  \caption{The ``double DIP'' model used in this work. Following prior
    work\cite{gandelsman_arxiv_2018},
  the outputs of both image-generating deep image priors (DIPs; as shown in Fig.~\ref{fig:dip}) are filtered by function $f_{1/2}$ to account
  for the partial band transfer of superimposed images. $f_{1/2}$ can be either a single-layer
  filter, or a shallow DIP network. }
  \label{fig:model}
\end{figure}

\begin{figure}[htbp]
  \centering
  \includegraphics[width=0.95\textwidth]{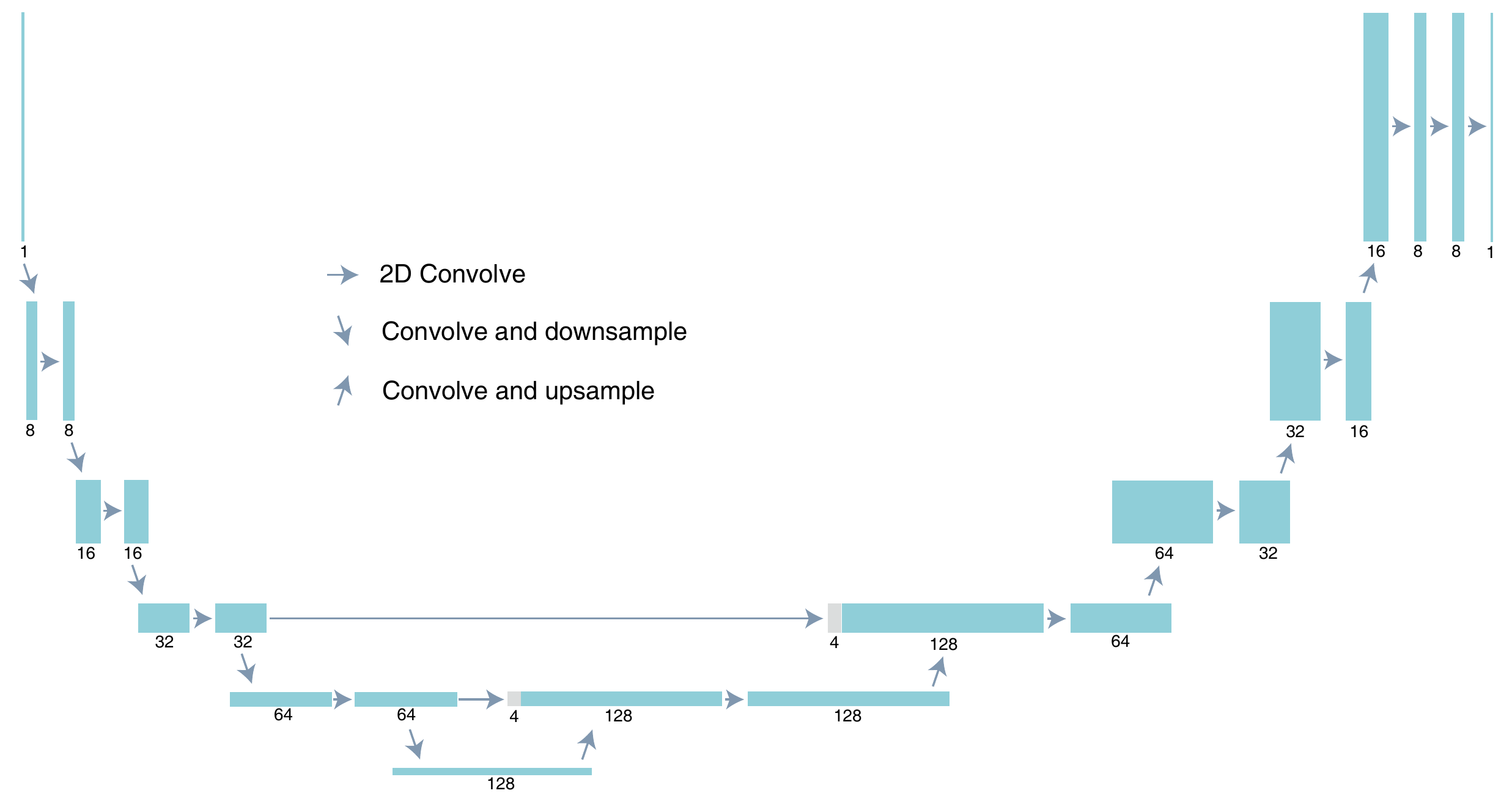}
  \caption{Architecture of a DIP network used in the DDIP model. Numbers underneath tensor blocks indicate
  the number of channels. }
  \label{fig:dip}
\end{figure}

The overall model structure of our modified DDIP is shown in Fig.~\ref{fig:model}. The two image-generating
DIPs, labeled DIP-1 and DIP-2, are of the same ``U-Net''-like
architecture \cite{ronneberger_arxiv_2015}, as shown in Fig.~\ref{fig:dip}. The kernel size used in all 2D
convolutional layers is $5\times 5$; an exception is the skip connections, where $1\times 1$ kernels
are used. The input/output numbers of channels of these convolutional layers are shown in the figure.
A leaky ReLU is used after
each 2D convolutional layer as the activation function.
The inputs to both DIPs, $z_1$ and $z_2$, are mono-channel tensors of random numbers that are
uniformly sampled between -0.5 and 0.5 and have the same height and
width as the original images.
The DIP that generates the constant weighting factor,
DIP-3, adopts the same architecture as DIP-1 and 2 except that the input and output numbers of channels are 2.
During training, DIP-1 and 2
learn to map $z_1$ and $z_2$ to $y_1$ and $y_2$ which are supposed to be the ``clean'' slice images.
For DIP-3, the values of the central pixels from both output channels are used as the blending weights
$\alpha_1$ and $\alpha_2$.
In \cite{gandelsman_arxiv_2018}, a linear combination is used
to synthesize the blended images $I_1$ and $I_2$ from generated images $y_1$ and $y_2$, \emph{i.e.},
$I_1 = \alpha_1 y_1 + (1 - \alpha_1)y_2$, and $I_2 = \alpha_2 y_1 + (1 - \alpha_2)y_2$. In our case, to
account for the band loss of the ghost features, we pass the images of the source of crosstalk through an additional
function $f_1$ or $f_2$, giving
\begin{align}
\begin{split}
   I_1 &= \alpha_1 y_1 + (1 - \alpha_1)f_1(y_2) \\
   I_2 &= \alpha_2 f_2(y_1) + (1 - \alpha_2)y_2.
   \label{eqn:forward}
\end{split}
\end{align}
We explored two types of choices for $f_1$ and $f_2$. Before
discussing the choices for these functions, one can see in Fig.~\ref{fig:auni_recons}(a)
and \ref{fig:zp_recons}(a) that the ghost images from more strongly scattering materials (\emph{e.g.}, gold)
appear like the high-pass filtered version of the real features. On the other hand, the more weakly
scattering materials (\emph{e.g.}, NiO) contribute to the crosstalk with a low-passed version of the real
features. Thus, one can define $f_1$ and $f_2$ as two single-kernel filtering functions, which can be
implemented through 2D convolution:
\begin{align}
\begin{split}
   f_1(x) &= x * k_1 \\
   f_2(x) &= x * k_2.
   \label{eqn:f1f2_filter}
\end{split}
\end{align}
Based on the appearance of the original images, $k_1$ and $k_2$ can be initialized to be a low-pass or
high-pass kernel. During training, their values are optimized along with the DIP parameters. For our
results to be shown in Section \ref{sec:results} where both cases are consisted of one slice with low-pass
crosstalk and another with high-pass crosstalk, we set $k_1$ to be a $7\times 7$ uniform filter,
and $k_2$ to be a $7\times 7$ kernel containing a 5-point Laplacian filter.

Using a single filtering kernel may not be able to capture the band loss at various spatial scales.
Therefore, a second way is to set $f_1$ and $f_2$ as another two shallow DIPs with downsampling
and skip connections. In our implementation, we used a 3-layer DIP
with the same kernel size as DIP-1, 2, and 3,
so that there are 3 downsampling/upsampling operations, each with a factor of 2.
However, the number of channels of intermediate tensors is always 1. Additionally, skip connections are
used at all 3 spatial scales in order to prevent the loss of high-frequency information. These shallow DIPs
are initialized using uniform random numbers, and the parameters are optimized along with the ``major''
DIPs during training.

With these, we can now formulate the loss function which contains a data mismatch term measuring the
difference between the synthesized images $I_{1/2}$ and the original images, $\ionetwont$. Additionally,
as indicated in \cite{gandelsman_arxiv_2018}, it is also essential to employ an exclusion loss which
penalizes the correlation of the spatial gradients of $y_1$ and $y_2$ at multiple spatial scales.
The values of $\alpha_1$ and $\alpha_2$ are also penalized for drifting away from 0.5 at the
first 100 epochs of the algorithm in order to stabilize their values against the random input and
network initialization. Thus, the full loss function (for a 2-slice separation task) is written as
\begin{equation}
  L = \sum_{i=1}^2\norm{I_i(\text{DIP-1,2,3},f_{1,2}) - I_i^0}^2 + \gammaexcl\sum_{j=1}^{5}\sum_{l\in\{x, y\}}{D_{j}(\nabla_l y_1)D_{j}(\nabla_l y_2)} + \chi_{[1, 100]}(k)\sum_{i=1}^2\norm{\alpha_i - 0.5}^2
  \label{eqn:loss}
\end{equation}
where $\gammaexcl$ is a constant weight of the exclusion loss term, $D_j$ is the downsampling function
that downsamples the image in its argument by a factor of $2^{j-1}$,  $\nabla_l y$ denotes the spatial
gradient of $y$ along direction $l$ (either $x$ or $y$), and $\chi_{[1, 100]}(k)$ is a step function of
epoch number $k$ that returns 1 when $k \leq 100$, and 0 otherwise.

Our model is trained on an HP Z8 G4 workstation with two Intel Xeon Silver 4108 CPUs and two NVIDIA Quadro
P4000 GPUs, although the model is run using only one GPU each time. PyTorch \cite{pytorch} is used for
automatic differentiation. The code is available on \url{https://github.com/mdw771/ddip4ptycho}.

\subsection{Beamline experiments}

The datasets used in both cases shown in Section \ref{sec:results} were acquired at the
Hard X-ray Nanoprobe beamline (3-ID) of the National Synchrotron Light
Source II at Brookhaven National Laboratory.

The first dataset
involves a synthetic sample, where Au nanoparticles and NiO particles are deposited on
both sides of a 10-\micron{}-thick Si wafer. We will hereafter refer to this dataset as the
Au/NiO dataset. The dataset was collected with a beam energy of 12 keV
and a transverse resolution $\deltares=7.3$ nm, which, according to Eq.~\ref{eqn:dof} gives $\zdof=2.8$ \micron{}.
The 10-\micron{} slice spacing is therefore about 3.6 times larger than $\zdof$.
The multislice reconstruction result of this dataset was published earlier in \cite{ozturk_optica_2018},
which can be referred for more experimental details. Similar to \cite{ozturk_optica_2018}, we assume
two slices in the sample, which respectively correspond to the Au layer and the NiO layer.

The second dataset, described here as the ZP/NiO dataset, also
involves a 2-slice sample that has been previously published \cite{huang_actacrysta_2019}.
In this case, Au zone plate structures and NiO particles are deposited on both sides of a
500-nm-thick silicon nitride membrane. The beam energy and transverse resolution on the first slice are
12 keV and 8.7 nm, giving $\zdof=3.9$ \micron{}. Hence, the slice spacing is just about 0.13 of
the DoF.

\section{Results}
\label{sec:results}

\subsection{Large-spacing separation for Au/NiO data}

\begin{figure}[htbp]
  \centering
  \includegraphics[width=0.95\textwidth]{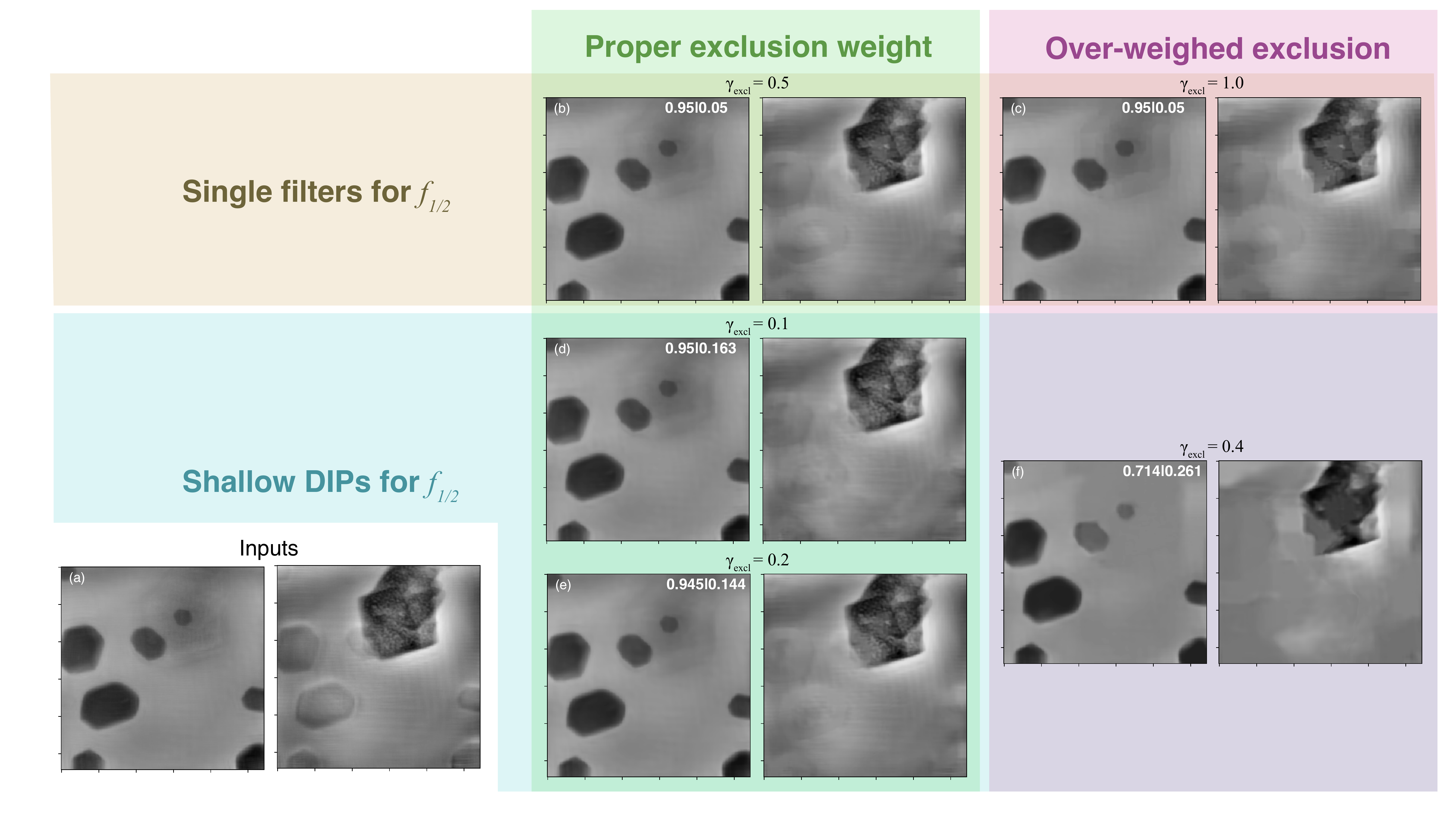}
  \caption{Input images (a) and separation results (b-f) of the Au/NiO dataset.
  The results were obtained with $f_{1/2}$ set to either single filters or shallow DIPs.
  For each case, several $\gammaexcl$ values were tested. Ghost features are effectively
  suppressed with a proper setting for $\gammaexcl$. However, when $\gammaexcl$ is too large,
  fine details of the features are smeared out.
  The final values of $\alpha_1|\alpha_2$ are indicated at the upper right corners
  of the corresponding subplots.
  The values of $\alpha$
  hold steady except when we use shallow DIPs for $f_{1/2}$ and over-weight the exclusion
  loss, in which case the ratio $\alpha_1 / \alpha_2$ decreases significantly.
  }
  \label{fig:auni_recons}
\end{figure}

The crosstalk-contaminated slice images of the Au/NiO dataset were reconstructed using an
adaptive momentum based algorithm in a tool we developed called``\textit{Adorym}'' \cite{du_arxiv_2020}.
The phase retrieval was initialized using Gaussian randoms, without using the XRF data.
Because the slice spacing is 3.6 times larger than $\zdof$, our multislice reconstruction
algorithm is able to provide reconstructions
of both slices with the ``true'' features of each slice resolved sharply, but they also exhibit
obvious ghost features due to the crosstalk. Next, we cropped a
$272\times 272$ pixel area that
has full probe overlap from each slice [Fig.~\ref{fig:auni_recons}(a)],
and passed the slices to DDIP as $\ionent$ and $\itwont$.

We performed 5 test runs with $f_{1/2}$ set to use either shallow DIPs or single filters for $f_{1/2}$,
and with different values of $\gammaexcl$. Each parameter combination
was run for 10000 epochs.
When using shallow DIPs for $f_{1/2}$, the peak GPU memory usage was 439 MB, and each run took
around 30 min to complete. The results are shown in Fig.~\ref{fig:auni_recons}(b-f), where
the final values of $\alpha_1$ and $\alpha_2$ are indicated at the top right corners of the
corresponding subplots as $\alpha_1|\alpha_2$. The dynamic range of all plots is set to
$[\mu - 4\sigma, \mu + 4\sigma]$, with $\mu$ and $\sigma$ being the image mean and standard deviation.

Since the ghost image on slice 1 of the NiO
particle (which is in fact on slice 2) is very blurry, it appears like a subtle change in the image background.
Under all tested parameter settings, DDIP barely affected the presence of this faint region.
This can be explained by the nature of deep image priors: as noted in \cite{gandelsman_arxiv_2018},
generative neural networks tend to generate images that have a smaller empirical entropy
across its local patches; in other words, the generated images tend to have
``strong internal self-similarity''. Since the ghost feature on slice 1 is very smooth, it is hard
for DIPs to exclude it from the generated image. However, the ghost features on slice 2 are
sharp and have a much higher variance. They make the local patches of the image more complicated and
more ``unlike'' to each other, so DIP tends to generate images that are free of these artifacts.
Therefore, the improvement of slice 2 are obvious. The effect on slice 2 is also
largely dependent on $\gammaexcl$ regardless whether
$f_{1/2}$ is set to use shallow DIPs or single filters. When using
single filters for $f_{1/2}$, the setting of
$\gammaexcl = 0.5$ can provide an apparent mitigation of the crosstalk coming from slice 1,
where the sharpness and contrast of the ghost Au particles are greatly reduced. Increasing
$\gammaexcl$ to 1.0 suppresses the ghost features even further, but it also starts to destroy
details in the ``true image'' of the NiO particle. In particular, the regions in the NiO
particle that overlap with ghost Au particles are severely smeared. Given such high values of
$\gammaexcl$, the correlation of gradients is over-penalized and the algorithm tends to reduce
the spatial gradient of slice 2 at the overlapping regions to 0, resulting in flattened areas.

Improved results are obtained when we switch $f_{1/2}$ to use shallow DIPs. In Fig.~\ref{fig:auni_recons}(d),
when $\gammaexcl = 0.1$, the crosstalk suppression on slice 2 is nearly as effective as
(b) with single filters and $\gammaexcl = 0.5$. Increasing $\gammaexcl$ to 0.2 slightly enhances
the suppression effect, surpassing the efficacy of Fig.~\ref{fig:auni_recons}(c) with single filters
and $\gammaexcl = 1.0$. Moreover, comparing (c) and (e) reveals that using shallow DIPs leads to
much better preserved high-frequency details in the NiO particle. This is an expected improvement,
as the multi-scale filtering with skip connections in the shallow DIPs better describes the
band loss of ghost features than single filters. If one increases $\gammaexcl$ further to 0.4,
however, the images would start to lose high-frequency details as well.

The final values of blending weights for all cases are composed of a large $\alpha_1$ and a
small $\alpha_2$. Based on Eq.~\ref{eqn:forward}, this indicates that $y_1$ contributes much
more than $f_1(y_2)$ does to $I_1$, while $y_2$ contributes more than $f_2(y_1)$ to $I_2$. This is a reasonable
trend as one would expect smaller contribution from the ghost
features than real features in a ``blended'' slice. However, we should not interpret the $\alpha$
values as the absolute intensities of the ghost or real features present in $I_1$ or $I_2$, since
the mean intensities of $y_1$, $y_2$, $f_1(y_2)$, and $f_2(y_1)$ can vary as well.
On the other hand, the $\alpha$ pair may be used as an indication of the fidelity of the
result. In Fig.~\ref{fig:auni_recons}(f) where the details of the features are obviously
undermined, the final value of $\alpha_1$ is much lower than other results with better preserved
features, while $\alpha_2$ is much higher. Since the algorithm always tries to minimize
the mismatch between $I_{1/2}$ and $\ionetwont$ where the latter is fixed, unusual $\alpha$
values point to unusual value ranges of the outputs of DIP-1/2 and $f_{1/2}$, implying that
the generated images might be highly aberrated.

\subsection{Small-spacing separation for ZP/NiO data}

\begin{figure}
  \centering
  \includegraphics[width=0.6\textwidth]{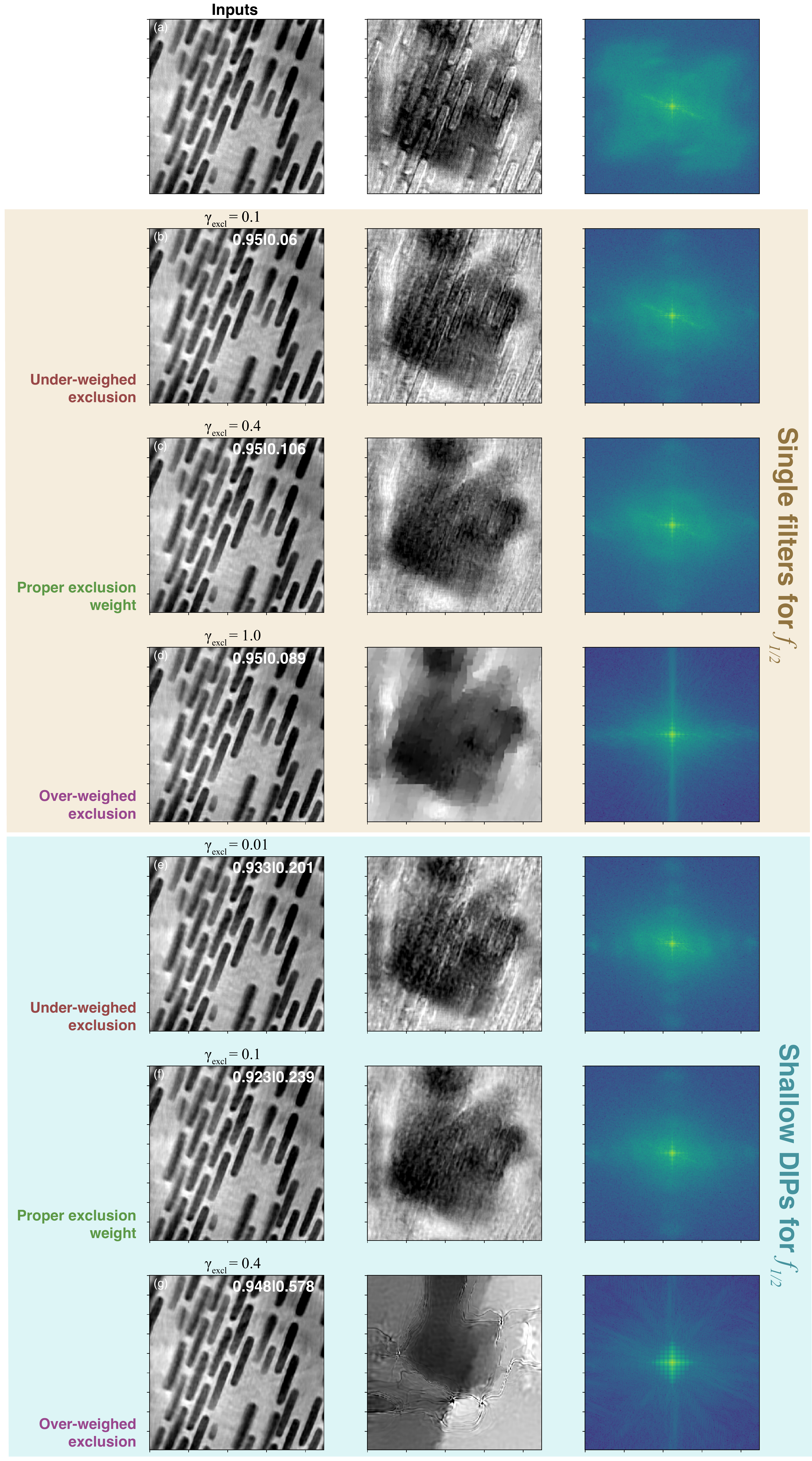}
  \caption{Input images (a) and separation results (b-g) of the ZP/NiO dataset.
  Like in the cases shown in Fig.~\ref{fig:auni_recons},
  the results were obtained with $f_{1/2}$ set to either single filters or shallow DIPs.
  The final values of $\alpha_1|\alpha_2$ are indicated at the upper right corners
  of the corresponding subplots.
  While the influence of $\gammaexcl$ on slice 1 is minimal, it greatly affects the
  balance between separation effectiveness and image resolution for slice 2.
  The rightmost column shows the normalized power spectra of slice 2. A slanted streak
  corresponding to the periodicity of the zone plate's ghost features can be seen obviously
  in the input image's spectrum. In the outputs of the DDIP (\emph{i.e.}, $y_2$), the spectrum density of this
  streak becomes much lower. Also, we again see
  that when we use shallow DIPs for $f_{1/2}$ and over-weigh the exclusion loss,
  a smaller $\alpha_1 / \alpha_2$ ratio is yielded. }
  \label{fig:zp_recons}
\end{figure}

The 500-nm slice spacing in the ZP/NiO dataset is only about 0.13 times the DoF.
As such, our attempt of reconstructing
both slices using random initial guesses yielded two slices that are largely undifferentiated. The superimposed features
on both slices are mixed with an almost identical ratio, and the band loss of ghost features is very small.
Images like this could hardly provide enough diversity of measurement in order to solve the BSS problem.
Therefore, it becomes essential to employ the XRF data as additional prior knowledge to the reconstruction
algorithm. As mentioned earlier, the slice images to be separated were obtained using the XRF-aided
method described in \cite{huang_actacrysta_2019}, where the XRF
map of Ni is used to reduce the contrast of NiO in the single-slice reconstruction, leaving the Au zone plate structure,
and the NiO-removed Au image and the re-sampled Ni XRF map are used as the initial guess for the first and second slice,
respectively, for the subsequent multislice ptychographic phase retrieval. Without dedicated parameter tuning
and algorithm search, standard phase retrieval could not provide well separated slices; instead, it yielded
the slice images shown in Fig.~\ref{fig:zp_recons}(a), where slice 2 is heavily affected by the ghost images
from the Au zone plate structures on slice 1. Our goal is to show that, even though XRF data have to be used,
DDIP can provide better separated images
based on this result, so that the excessive amount of phase retrieval parameter tuning may be avoided.

We again tested several $\gammaexcl$ values with $f_{1/2}$ using shallow DIPs or single filters. 10000 epochs
are run for each case. When using shallow DIPs, the peak memory usage is 1130 MB, and it took 37 min to complete
the training. On the other hand, when using single filters, the total walltime becomes 31 min, though the peak
memory usage did not change significantly since the parameter size of the shallow DIPs is rather small compared
to the major DIPs.
The results are shown in Fig.~\ref{fig:zp_recons}(b-h).
Similar to what was observed with the Au/Ni dataset, the crosstalk does not significantly affect slice 1, but results in obvious ghost
images on slice 2 due to the strong scattering of Au. For single filters, $\gammaexcl = 0.4$ [Fig.~\ref{fig:zp_recons}(c)]
gives the best balance between crosstalk suppression and feature fidelity. Using a lower $\gammaexcl$ of 0.1
leaves a lot of residual ghost image features, while a higher value of 1.0 results in a blocky appearance of the
recovered slice 2. When using shallow DIPs, the optimal $\gammaexcl$ is found around 0.1. If $\gammaexcl$ is
set too high, the fidelity of $y_2$ is dramatically lost, which is accompanied by a much larger $\alpha_2$.

Since the Au zone plate structures are well aligned in the same direction, we can analyze the power spectra
of the outcome $y_2$ to evaluate the effectiveness of crosstalk suppression. These power spectra are normalized
by the integrated energy, and plotted on a logarithmic scale. In the power spectra of the original image
shown in
Fig.~\ref{fig:zp_recons}(a), one can clearly observe a slanted streak that represents the periodicity
of the zone plate ghost features. For $\gammaexcl = 0.4$ when using single filters and $\gammaexcl = 0.1$
when using shallow DIPs, the streak becomes barely visible. Further increasing $\gammaexcl$ in both cases
cause energy to concentrate in the low-frequency region, associated with the smeared appearance of (d) and
(g).

\section{Discussion}

\begin{figure}
  \centering
  \includegraphics[width=0.95\textwidth]{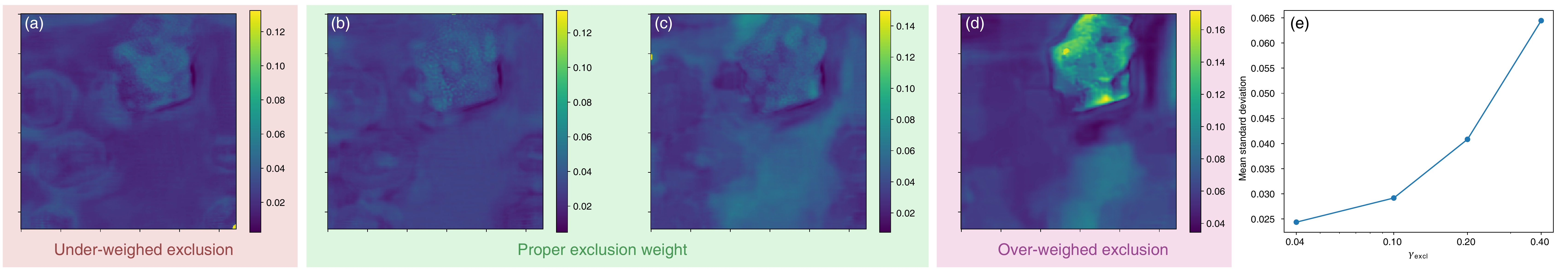}
  \caption{(a-d) Standard deviation maps of output slice 2 ($y_2$) of the Au/Ni dataset, calculated
  from 20 independent runs, for $\gammaexcl = 0.04, 0.1, 0.2, 0.4$.
  (e) plots the mean standard deviation against $\gammaexcl$. These standard deviation values measure
  the uncertainty of DDIP, as the input vectors to the DIPs are randomly initialized for each run.
  Larger $\gammaexcl$ results in larger uncertainty.}
  \label{fig:auni_uncertainty}
\end{figure}

We have demonstrated the crosstalk separation capability of our modified DDIP model in two cases, one
with slice images reconstructed without using XRF data, the other
reconstructed using the aid of XRF data
but without fine tuning of phase retrieval parameters. In practice, one problem of concern might be
the reproducibility of the algorithm due to its inherent uncertainty, which is contributed by the randomness
of input vectors $z_1$, $z_2$, $z_3$ and the random initialization of network parameters. In our
experience, this uncertainty is associated with the value of $\gammaexcl$, so we conducted a test
to evaluate the change of result distribution with $\gammaexcl$. On the Au/Ni dataset, we ran a series
of DDIP separations using $\gammaexcl = 0.04, 0.1, 0.2, 0.4$, each run for 20 times. For the results
of each $\gammaexcl$, the standard deviation over the 20 runs at each pixel position is shown in
Fig.~\ref{fig:auni_uncertainty}(a-d). The averages of these standard deviation maps are plotted
in Fig.~\ref{fig:auni_uncertainty}(e), which clearly show an increasing trend. Referring back to
Fig.~\ref{fig:auni_recons}, the optimal result sing shallow DIPs is obtained with $\gammaexcl = 0.2$,
where the image mean is 0.68, but the uncertainty standard deviation is only around 0.04. In practice,
one can also perform multiple runs and use the average $y_1$ and $y_2$ as the final results, so as to
further decrease the uncertainty. Other than the detailed variation of the separated images, it is
also possible for DDIP to undergo ``slice confusion'': since the inputs to the generating DIPs are
purely random, they do not inform DDIP that $y_1$ should correspond to real features
on slice 1, and vice versa for $y_2$. If DDIP is confused about the slice arrangement,
it may tend to generate the real,
solid Au particles, which should lie on slice 1, on $y_2$ instead. According to
Eq.~\ref{eqn:forward}, these Au particles will be filtered
by $f_1$ to form $I_1$, which is unphysical; the same would apply to $I_2$. However, with a reasonable
$\gammaexcl$, this is very unlikely to happen: following the example above, if the real Au particles
appear on $y_2$, then they will appear unfiltered on $I_2$; yet, on $\itwont$ these particles are highpass
filtered, and this leads to high mismatch loss which in unfavored. Therefore, the band loss of feature
blending and our unsymmetrical use of $f_{1/2}$ in the forward model drive the DDIP towards the correct
slice arrangement. In our uncertainty test, we did not see slice confusion in all of our 80 separation
results.

Both results shown in Section \ref{sec:results} involve 2 slices. In practice, multislice ptychography
may be used to reconstruct 3 slices or more, and mutual crosstalk may involve more than 2 slices.
In that case, one may add more DIPs, so that the number of image-generating DIPs matches the number
of mutually crosstalking slices $N$. Meanwhile, the input and output channels of the weight-generating DIP
may be increased to $N$, and the forward model of Eq.~\ref{eqn:forward} may be expanded
to $N$ equations, constituting an $N\times N$ mixing matrix. Using too many DIPs will unavoidably
impair the efficiency of the algorithm. However, in x-ray microscopy, the number of slices is typically
small due to the large DoF of X rays. Making the DDIP method more efficient for many-slice problems
is a future direction to explore.

\section{Conclusion}

Using a modified double-DIP architecture, we demonstrated the use of deep neural networks in mitigating
the crosstalk artifacts of multislice ptychography phase retrieval. When the slice spacing is large
(many multiples of the DoF), phase retrieval from scratch can provide slice reconstructions that are
distinct from each other but affected by crosstalk,
while post-processing using DDIP may suppress or remove the crosstalk on each slice. Combining multislice
phase retrieval and DDIP can yield good reconstructions without XRF data in this case. When the slice
spacing is small, phase retrieval may need the aid of XRF data in order to generate distinguishable
slice images, and the retrieved images may still contain crosstalk artifacts without dedicated parameter
tuning. One can also use DDIP in this case to suppress the crosstalk, so that one no longer has to
spend time searching for the best values of phase retrieval hyperparameters. In order to account for the band loss of crosstalking
features in a slice image, we pass the them through a filtering function in our forward model. The
filtering function can take the form of either a single convolutional filter or a shallow DIP. While
the former is faster, the latter can often provide results with better preserved details. We expect that
the findings will help improve the adaptability of multislice ptychography in imaging thick samples
beyond the DoF limit.

\section*{Acknowledgement}

This research used resources of the Advanced Photon Source (APS), a
U.S. Department of Energy (DOE) Office of Science User Facility
operated for the DOE Office of Science by Argonne National Laboratory
under Contract No. DE-AC02-06CH11357.
It also used 3ID of the National Synchrotron Light Source II,
a U.S. Department of Energy (DOE) Office of Science User Facility
operated for the DOE Office of Science by Brookhaven National
Laboratory under Contract No. DE-SC0012704.
We thank the Argonne Laboratory Directed Research
and Development for support under grant 2019-0441.
We also thank the National Institute of Mental
Health, National Institutes of Health, for support under grant R01
MH115265. We appreciate the authors of the ``double-DIP''
paper (Gandelsman \textit{et al.} \cite{gandelsman_arxiv_2018}) for sharing their code, which we
adapted and modified for this work.

\section*{Disclosure}

The authors declare no conflicts of interest.

\bibliographystyle{unsrt}
\bibliography{mybib}

\begin{thebibliography}{10}

\bibitem{faulkner_prl_2004}
H~M~L Faulkner and J~Rodenburg.
\newblock Movable aperture lensless transmission microscopy: A novel phase
  retrieval algorithm.
\newblock {\em Physical Review Letters}, 93(2):023903, July 2004.

\bibitem{maiden_josaa_2012}
A~M Maiden, M~J Humphry, and J~M Rodenburg.
\newblock Ptychographic transmission microscopy in three dimensions using a
  multi-slice approach.
\newblock {\em Journal of the Optical Society of America A}, 29(8):1606--1614,
  August 2012.

\bibitem{tsai_optexp_2016b}
Esther H~R Tsai, Ivan Usov, Ana Diaz, Andreas Menzel, and Manuel
  Guizar-Sicairos.
\newblock X-ray ptychography with extended depth of field.
\newblock {\em Optics Express}, 24(25):29089--29108, 2016.

\bibitem{born_optics}
Max Born, Emil Wolf, A.~B. Bhatia, P.~C. Clemmow, D.~Gabor, A.~R. Stokes, A.~M.
  Taylor, P.~A. Wayman, and W.~L. Wilcock.
\newblock {\em Scattering from inhomogeneous media}, pages 695--734.
\newblock Cambridge University Press, 7 edition, 1999.

\bibitem{gilles_optica_2018}
M~A Gilles, Y~S~G Nashed, M~Du, C~Jacobsen, and S~M Wild.
\newblock {3D} x-ray imaging of continuous objects beyond the depth of focus
  limit.
\newblock {\em Optica}, 5:1078--1085, 2018.

\bibitem{ozturk_optica_2018}
Hande {\"O}zt{\"u}rk, Hanfei Yan, Yan He, Mingyuan Ge, Zhihua Dong, Meifeng
  Lin, Evgeny Nazaretski, Ian~K Robinson, Yong~S Chu, and Xiaojing Huang.
\newblock Multi-slice ptychography with large numerical aperture multilayer
  {Laue} lenses.
\newblock {\em Optica}, 5(5):601--607, May 2018.

\bibitem{du_ultramic_2018}
Ming Du and Chris Jacobsen.
\newblock Relative merits and limiting factors for x-ray and electron
  microscopy of thick, hydrated organic materials.
\newblock {\em Ultramicroscopy}, 184:293--309, 2018.

\bibitem{dierolf_nature_2010}
Martin Dierolf, Andreas Menzel, Pierre Thibault, Philipp Schneider, Cameron~M
  Kewish, Roger Wepf, Oliver Bunk, and Franz Pfeiffer.
\newblock Ptychographic x-ray computed tomography at the nanoscale.
\newblock {\em Nature}, 467(7314):436--439, September 2010.

\bibitem{vandenbroek_prl_2012}
Wouter Van~den Broek and Christoph~T Koch.
\newblock Method for retrieval of the three-dimensional object potential by
  inversion of dynamical electron scattering.
\newblock {\em Physical Review Letters}, 109(24):245502, December 2012.

\bibitem{kamilov_optica_2015}
Ulugbek~S Kamilov, Ioannis~N Papadopoulos, Morteza~H Shoreh, Alexandre Goy,
  Cedric Vonesch, Michael Unser, and Demetri Psaltis.
\newblock Learning approach to optical tomography.
\newblock {\em Optica}, 2(6):517--522, 2015.

\bibitem{li_scirep_2018}
Peng Li and Andrew Maiden.
\newblock Multi-slice ptychographic tomography.
\newblock {\em Scientific Reports}, 8:2049, January 2018.

\bibitem{du_sciadv_2020}
Ming Du, Youssef S~G Nashed, Saugat Kandel, Dog{\u a} G{\"u}rsoy, and Chris
  Jacobsen.
\newblock Three dimensions, two microscopes, one code: automatic
  differentiation for x-ray nanotomography beyond the depth of focus limit.
\newblock {\em Science Advances}, 6:eaay3700, March 2020.

\bibitem{huang_actacrysta_2019}
Xiaojing Huang, Hanfei Yan, Yan He, Mingyuan Ge, Hande \"{O}ztürk, Yao-Lung~L.
  Fang, Sungsoo Ha, Meifeng Lin, Ming Lu, Evgeny Nazaretski, Ian~K. Robinson,
  and Yong~S. Chu.
\newblock Resolving 500 nm axial separation by multi‐slice x‐ray
  ptychography.
\newblock {\em Acta Crystallographica Section A}, 75(2):336--341, 2019.

\bibitem{cao_ieeetrans_1996}
Xi-Ren Cao and Ruey-Wen Liu.
\newblock General approach to blind source separation.
\newblock {\em IEEE Transactions on Signal Processing}, 44(3):562--571, 1996.

\bibitem{ulyanov_arxiv_2018}
Dmitry Ulyanov, Andrea Vedaldi, and Victor Lempitsky.
\newblock Deep image prior.
\newblock {\em 2018 IEEE/CVF Conference on Computer Vision and Pattern
  Recognition}, pages 9446--9454, 2018.

\bibitem{gandelsman_arxiv_2018}
Yossi Gandelsman, Assaf Shocher, and Michal Irani.
\newblock {``Double-DIP''}: Unsupervised image decomposition via coupled
  deep-image-priors.
\newblock arXiv:1812.00467, 2018.

\bibitem{ronneberger_arxiv_2015}
Olaf Ronneberger, Philipp Fischer, and Thomas Brox.
\newblock {U-Net}: Convolutional networks for biomedical image segmentation.
\newblock arXiv:1505.04597, 2015.

\bibitem{pytorch}
Adam Paszke, Sam Gross, Francisco Massa, Adam Lerer, James Bradbury, Gregory
  Chanan, Trevor Killeen, Zeming Lin, Natalia Gimelshein, Luca Antiga, Alban
  Desmaison, Andreas Kopf, Edward Yang, Zachary DeVito, Martin Raison, Alykhan
  Tejani, Sasank Chilamkurthy, Benoit Steiner, Lu~Fang, Junjie Bai, and Soumith
  Chintala.
\newblock Pytorch: An imperative style, high-performance deep learning library.
\newblock In H.~Wallach, H.~Larochelle, A.~Beygelzimer, F.~d\textquotesingle
  Alch\'{e}-Buc, E.~Fox, and R.~Garnett, editors, {\em Advances in Neural
  Information Processing Systems 32}, pages 8024--8035. Curran Associates,
  Inc., 2019.

\bibitem{du_arxiv_2020}
Ming Du, Saugat Kandel, Junjing Deng, Xiaojing Huang, Arnaud Demortiere,
  Tuan~Tu Nguyen, Remi Tucoulou, Vincent~De Andrade, Qiaoling Jin, and Chris
  Jacobsen.
\newblock Adorym: A multi-platform generic x-ray image reconstruction framework
  based on automatic differentiation.
\newblock arXiv:2012.12686, 2020.

\end{thebibliography}

\end{document}